\documentclass[10pt,conference]{IEEEtran}
\IEEEoverridecommandlockouts
\usepackage{cite}
\usepackage{amsmath,amssymb,amsfonts}
\usepackage{algorithmic}
\usepackage{graphicx}
\usepackage{textcomp}
\usepackage{xcolor}
\def\BibTeX{{\rm B\kern-.05em{\sc i\kern-.025em b}\kern-.08em
    T\kern-.1667em\lower.7ex\hbox{E}\kern-.125emX}}
\begin{document}

\title{Towards Understanding Machine Learning Testing in Practice}

\author{\IEEEauthorblockN{Arumoy Shome}
\IEEEauthorblockA{\textit{Software Engineering Research Group} \\
\textit{Delft University of Technology}\\
Delft, Netherlands \\
a.shome@tudelft.nl}
\and
\IEEEauthorblockN{Lu{\'\i}s Cruz}
\IEEEauthorblockA{\textit{Software Engineering Research Group} \\
\textit{Delft University of Technology}\\
Delft, Netherlands \\
l.cruz@tudelft.nl}
\and
\IEEEauthorblockN{Arie van Deursen}
\IEEEauthorblockA{\textit{Software Engineering Research Group} \\
\textit{Delft University of Technology}\\
Delft, Netherlands \\
arie.vandeursen@tudelft.nl}
}

\maketitle

\begin{abstract}

  Visualisations drive all aspects of the Machine Learning (ML)
  Development Cycle but remain a vastly untapped resource by the
  research community. ML testing is a highly interactive and cognitive
  process which demands a human-in-the-loop approach. Besides writing
  tests for the code base, bulk of the evaluation requires application
  of domain expertise to generate and interpret visualisations. To
  gain a deeper insight into the process of testing ML systems, we
  propose to study visualisations of ML pipelines by mining Jupyter
  notebooks. We propose a two prong approach in conducting the
  analysis. First, gather general insights and trends using
  a qualitative study of a smaller sample of notebooks. And then use
  the knowledge gained from the qualitative study to design an
  empirical study using a larger sample of notebooks. Computational
  notebooks provide a rich source of information in three
  formats---text, code and images. We hope to utilise existing work in
  image analysis and Natural Language Processing for text and code, to
  analyse the information present in notebooks. We hope to gain a new
  perspective into program comprehension and debugging in the context
  of ML testing.

\end{abstract}


\section{Introduction}
\label{sec:intro}

Visualisations are the ``bread and butter'' of a data scientist since
they drive all aspects of the Machine Learning Development
Cycle. Visualisations are used as early as the data exploration phase
to understand the underlying dataset and gain insights prior to
training. During the training phase, visualisations are used to
evaluate and compare the predictive performance of various ML
models. Once a model is selected, visualisations aid practitioners in
testing black-box ML systems for non-functional properties such as
fairness and explainability \cite{zhang2020machine}. Finally, once
a model is deployed, visualisations are used to continually monitor
its health and automatically trigger a new training cycle if the
performance degrades due to data drifts \cite{breck2019data}.



Computational Notebooks have become wide-adopted by the data science
community to develop ML pipelines. Computational notebooks are
a perfect fit for the data science workflow as they allow
practitioners to interweave text, code and visualisations in a single
cohesive document. Although there is an abundance of publicly
available computational notebooks, they still remain a vastly untapped
resource by the research community.

Testing ML systems is a highly interactive and cognitive process which
demands a human-in-the-loop approach. In addition to writing tests for
the code base, bulk of the evaluation requires application of domain
expertise to generate and interpret visualisations. ML testing is
a relatively new field of research. As such, many of the contributions
work in an experimental setting, however their feasibility in a more
practical environment remains unclear.

We propose a novel approach to understanding the current challenges of
ML testing in practice. To gain a deeper insight into the process of
ML testing itself, we propose to study the visualisations generated
for ML pipelines by mining Jupyter notebooks in the wild.

\section{Methodology}
\label{sec:method}

We propose to collect Jupyter notebooks from \emph{Kaggle}---a popular
online repository for data science and ML computational
notebooks\footnote{https://kaggle.com}. Kaggle provides a stable API
to download notebooks based on filters specified by the user. 

This allows us to adopt different search strategies based on the goal
of the study. To gain a general perspective on ML testing, we can mine
notebooks from popular Kaggle competitions. Alternatively, we may
choose to focus on a single functional or non-functional test
property. For instance, to understand how practitioners test for
fairness in ML systems, we can collect notebooks that are associated
with data science competitions focusing on fairness or datasets that
have been cited by prior scientific contributions in fairness testing. 

We only consider notebooks written in
Python\footnote{https://python.org} due to its popularity and rich
ecosystem of data science packages. As an additional measure of
quality, we only consider notebooks that are fully reproducible in
a containerised Docker environment and utilise stable and well tested
python packages\footnote{Jupyter provides a Docker image containing
all popular packages
https://hub.docker.com/r/jupyter/datascience-notebook}.

Jupyter notebooks contain a mix of plain-text (written in Markdown, a
popular markup
language\footnote{https://daringfireball.net/projects/markdown/}) and
code cells. Since notebook cells are represented internally as JSON
fields\footnote{https://ipython.org/ipython-doc/3/notebook/nbformat.html},
Jupyter notebooks are machine parsable thus allowing us to separate the
plain-text and code cells into individual files.


We wish to conduct the analysis in two phases. First, a qualitative
study using a smaller sample of notebooks (for instance, 5\% of the
top voted notebooks from the top 5 competitions or datasets) to gather
general insights and trends. And next, utilise our knowledge from the
qualitative study to design an empirical study using a larger sample
of notebooks.

\section{Challenges}
\label{sec:challenge}

The data collection process of downloading a sufficiently large sample
of Jupyter notebooks poses a significant engineering challenge. This
requires a working knowledge of web technologies to programmatically
download notebooks using the Kaggle API. Furthermore, reproducing all
notebooks in a Docker container can be a resource and time exhaustive
process and thus needs to be efficiently parallelised across the
resources of the computing device.

Notebooks are notoriously well known for not adhering to software
engineering standards and best practices. In addition to external
packages, notebooks also require the associated dataset(s) in order to
be reproduced. Prior work in mining Jupyter notebooks is
limited. However there are some relevant contributions that we wish to
adapt for our study. Specifically, \emph{Pimentel et
al} \cite{pimentel2019large} and \emph{Quaranta et
al} \cite{quaranta2021kgtorrent} mine a large quantity of Jupyter
notebooks which can aid with the data collection pipeline for this
study and provide helpful guidelines on reproducing computational
notebooks.

We also expect significant challenges in isolating and extracting the
code cells that generate visualisations. \emph{Bavishi et
al} \cite{bavishi2021vizsmith} mine Jupyter notebooks to create an
automated tool that suggests visualisation code for a given dataset
and text prompt from the user. We wish to adapt their methodology for
isolating visualisation code cells, to collect the images for our
study.

\section{Expected Outcomes}
\label{sec:contrib}

Jupyter notebooks present a rich source of information in three
different formats---text, code and images. We hope to gain insights
into the ML testing process by mining each source of information
separately. To the best of our knowledge, this has not been attempted
before.

We hope to utilise the existing literature on image analysis to derive
interesting insights from the visualisations collected in the
study \cite{nixon2019feature,parker2010algorithms}. Additionally, this
study also presents us with an opportunity to utilise the
state-of-the-art developments in Natural Language Processing (NLP)
both for plain-text and
code \cite{devanbu2020deep,zan2022neural,allamanis2018survey}. By
combining all three information sources, we hope to gain a new
perspective into program comprehension and debugging in the context of
ML testing.

\section{Outcomes Beyond ML Testing}

We also see opportunities to make scientific contributions that go
beyond ML testing. In particular, we hope to gain interesting insights
into reproducibility and maintainability of computational notebooks
from our data collection and pre-processing pipeline.

Our systematic methodology for obtaining, parsing and analysing
computational notebooks may be packaged into an automated analytics
tools. Given a notebook as input, the tool should produce meaningful
insights using all three data sources. With this tool, we wish to aid
future researchers who want to mine computational notebooks.

Prior studies have been conducted to understand ML pipelines and
formalise their development
process \cite{sambasivan2021everyone,sculley2015hidden}. However the
data collected from these studies are not openly accessible due to
confidentially restrictions. Our methodology of mining computational
notebooks can also be used to validate prior claims and provide
empirical evidence which is reproducible and publicly available.

We also see links to explainability in AI systems. Specifically, we
hope to gain insights into what tools and techniques practitioners are
using to understand black-box models and how they vary as the
complexity of the underlying data and ML model changes.

\bibliographystyle{IEEEtran}
\bibliography{report}

\end{document}